\documentclass[preprint]{aastex}

\usepackage{graphicx}

\makeatletter

\begin{document}

\title{Scattered Light from Dust in the Cavity of the V4046 Sgr Transition Disk}

\author
{Valerie A. Rapson\altaffilmark{1},
 Joel H. Kastner\altaffilmark{1},
 Sean M. Andrews\altaffilmark{2}, 
  Dean C. Hines\altaffilmark{3},
 Bruce Macintosh\altaffilmark{4},
 Max Millar-Blanchaer\altaffilmark{5},
 Motohide Tamura\altaffilmark{6}
}

\email{var5998@rit.edu}
\altaffiltext{1}{School of Physics and Astronomy, Rochester Institute of Technology, 1 Lomb Memorial Drive, Rochester, NY 14623-5603, USA}
\altaffiltext{2}{Harvard-Smithsonian Center for Astrophysics, 60 Garden Street, Cambridge, MA 02138, USA}
\altaffiltext{3}{Space Telescope Science Institute, Baltimore, MD, USA}
\altaffiltext{4}{Physics Department, Stanford University, Stanford, CA 94305, USA}
\altaffiltext{5}{Department of Astronomy and Astrophysics, University of Toronto, ON, M5S 3H4, Canada }
\altaffiltext{6}{National Astronomical Observatory of Japan, Mitaka, Tokyo 181-8588, Japan}

\begin{abstract}

We report the presence of scattered light from dust grains located in the giant planet formation region of the circumbinary disk orbiting the $\sim$20-Myr-old close ($\sim$0.045 AU separation) binary system V4046 Sgr AB based on observations with the new Gemini Planet Imager (GPI) instrument. These GPI images probe to within $\sim$7 AU of the central binary with linear spatial resolution of $\sim$3 AU, and are thereby capable of revealing dust disk structure within a region corresponding to the giant planets in our solar system. The GPI imaging reveals a relatively narrow (FWHM $\sim$10 AU) ring of polarized near-infrared flux whose brightness peaks at $\sim$14 AU.
This $\sim$14 AU radius ring is surrounded by a fainter outer halo of scattered light extending to $\sim$45 AU, which coincides with previously detected mm-wave thermal dust emission. The presence of small grains that efficiently scatter starlight well inside the mm-wavelength disk cavity supports current models of planet formation that suggest planet-disk interactions can generate pressure traps that impose strong radial variations in the particle size distribution throughout the disk. 
\end{abstract}

\keywords{circumstellar matter, polarization, stars: pre-main sequence, stars: individual (V4046 Sgr)}

\section{Introduction}

According to the core accretion model of gas giant planet formation, such planets are formed in circumstellar disks via agglomeration of dust particles into a planetary core, followed by accretion of disk dust and gas \citep[e.g.][and references therein]{Helled2013, Kley2012}. Disk material co-orbiting with a sufficiently massive planet is transported outward via deposition of angular momentum onto the disk, thus opening a large gap \citep[e.g.,][]{Bryden1999,Dong2014}. Density waves from planet-disk interactions generate pressure maxima in the disk that manifest themselves as ring-like or spiral disk structures characterized by sharp radial gradients in both surface density and particle size \citep{Pinilla2012}. Such dust ring structures have been observed around nearby, young (age $<$10 Myr), low-mass stars in optical and near-infrared scattered light images and sub-mm thermal emission \citep{Hughes2007,Andrews2009,Isella2010,Andrews2011,Garufi2013}. These same systems show depletions of cm- and mm-sized dust grains within the central regions of their disks. Such central disk ``holes" have been interpreted as evidence of formation of giant planets with orbital semimajor axes of $\lesssim$ 30 AU, i.e., well interior to the sub-mm bright rings \citep{Pinilla2012,Zhu2012,Garufi2013,ZHU2014,Owen2014}.

Thanks to its proximity \citep[D$\sim$73 pc;][]{torres2008} and advanced age \citep[$\sim$23 Myr;][]{Mamajek2014}, the circumbinary disk orbiting the close (2.4 day period) binary system V4046 Sgr AB represents an excellent subject for the study of such late-stage planet-building processes. The V4046 Sgr AB system consists of two nearly equal mass components \citep[0.9 and 0.85 M$_{\odot};$][]{Rosenfeld2012} with spectral types K5Ve and K7Ve \citep{Stemples2004} separated by only $\sim$9 $R_\odot$ (0.045 AU). Since the recognition that V4046 Sgr AB possesses a large circumbinary dust mass \citep{Jensen1997} and is orbited by (and actively accreting from) a gaseous disk \citep{Stemples2004,Gunther2006,Kastner2008}, this system has been extensively studied from X-ray to sub-mm wavelengths \citep{Rodriguez2010,Donati2011,Oberg2011,Argiroffi2012,Rosenfeld2012,Rosenfeld2013, Kastner2014}. Radio interferometric imaging of the V4046 Sgr disk in CO emission reveals that the circumbinary molecular disk is inclined at 33.5$^\circ$ \citep{Rosenfeld2012,Rodriguez2010} and extends to $\sim$350 AU \citep{Rodriguez2010}, with an estimated total gas+dust mass of $\sim$0.1 $M_\odot$ \citep{Rosenfeld2013}. Such a large and massive disk is unexpected, given that the V4046 Sgr system is a factor $\sim$10 older than the vast majority of known, actively accreting stars with circumstellar disks \citep[e.g.,][]{Ingleby2014}. The V4046 Sgr disk also displays more compact, distinctly ring-like mm continuum emission, suggesting a depletion of mm-sized dust interior to $\sim$29 AU due to dust particle growth and particle size segregation associated with recent or ongoing planet formation \citep{Rosenfeld2013}.

Polarimetric near-infrared imaging of light scattered off dust particles is useful for determining the radial and azimuthal distribution of micron-sized dust within the planet forming regions of circumstellar disks \citep[e.g.][]{Avenhaus2014,Hashimoto2011}. These observations serve as a powerful complement to mm-wave interferometric imaging in establishing radial gradients in dust particle size \citep[e.g.][]{Dong2012}. Here, we present coronagraphic/polarimetric images of the circumbinary disk around V4046 Sgr obtained with the Gemini Planet Imager \citep[GPI;][]{Macintosh2008,Macintosh2014} on Gemini South. These images probe closer to the central star(s) than previously achieved for a gas-rich, protoplanetary disk (R$\sim$7 AU) with unprecedented linear spatial resolution of $\sim$3 AU, and are thereby capable of revealing the dust disk structure within a region corresponding to the giant planets in our solar system. 

\section{Observations and Data Reduction} 

Early Science coronagraphic/polarimetric images of V4046 Sgr were obtained with Gemini/GPI through J (1.24 $\mu$m) and K2 (2.27 $\mu$m) band filters and 0.184 and 0.306$''$ diameter coronagraphic spots on April 23 and 24, 2014, respectively.  Four sets of J (K2) band images were obtained at waveplate position angles of 0, 22, 45 and 68$^\circ$ with exposure times of 30 s (60 s) through airmasses ranging from 1.24-1.97 (1.00-1.01) and DIMM seeing $\sim$0.65" ($\sim$0.60").

Exposures of 60 s were initially attempted in the J band, but saturated the detector. This saturation may have been due to a slight mispointing, causing starlight to leak out the west side of the coronagraph and saturate the detector.  Although GPI's extreme adaptive optics delivers very low total wavefront error, vibrations from the cryocoolers and a small fixed focus offset between the coronagraph focal plane mask and the science focal plane degraded the angular resolution of Early Science data to  $\sim0.05"$ ($\sim$ 3 AU) at J band. At the time of observation, V4046 Sgr (m$_{I}$= 9.11 mag) was the faintest target successfully observed with GPI and hence served as a test of the useful domain of the GPI instrument in coronagraphic/polarimetric mode.

 Images in each filter were reduced and combined using the GPI pipeline v1.2.1 \citep{Maire2010, Perrin2014_pipe}, following methods similar to those outlined in \citet{Perrin2014}. These basic reduction steps include image background subtraction, removal of correlated noise due to detector readout electronics and microphonics, and interpolation over bad pixels. Calibration spot grids, which define the location of each polarization spot pair produced by the lenslet array, were used to extract the data from each raw image and produce a pair of orthogonally polarized images. Satellite spots on each J-band image were used to determine the location of the (unresolved) binary star behind the coronagraph. The binary was insufficiently bright in the K2 images for accurate determination of its location behind the occulting spot via this technique, so its location was assumed to be at the center of the apparent coronagraph spot in each image.  To avoid positive bias in the polarized intensity image, the radial and tangential Stokes parameters P$_{\parallel}$ and P$_\perp$ \citep{Avenhaus2014b} were computed via the pipeline; we assume that all polarized flux is in the tangential component. Procedures for subtraction of the total intensity PSF for extended objects in GPI's polarization mode are still under development. We therefore focus our analysis on the P$_{\perp}$ images.

 \section{Results}
Figure 1 shows the total intensity, P$_\perp$, and scaled P$_\perp$ images of the disk around V4046 Sgr at J and K2. The orientation of the polarization (pseudo-)vectors are elliptically symmetric around the occulted central binary and the polarization fraction is much greater than the instrumental polarization of $\sim$0.4\% \citep{Wiktorowicz2014}. These results are as expected for scattering of starlight off circumstellar dust. The P$_\perp$  and scaled P$_\perp$ images in Figure 1 show a relatively narrow, bright central ring that peaks in brightness at $\sim$14 AU and is surrounded by a fainter, outer halo detected at $\ge$2$\sigma$ out to $\sim$45 AU. The scaled P$_\perp$ images, which account for dilution of incident starlight  \citep[e.g.][]{Garufi2014} highlight this ring/halo structure. The structure is detected at both J and K2, but is most clearly seen in the former because this shorter near-infrared wavelength probes dust that --- in addition to being illuminated by a brighter incident stellar radiation field --- likely has a larger scattering efficiency. The inner ring also shows radial dark features at J, which could be shadowing from dust within or interior to the bright ring.

\begin{figure}[htbp]
\begin{center}
\includegraphics[width=2in]{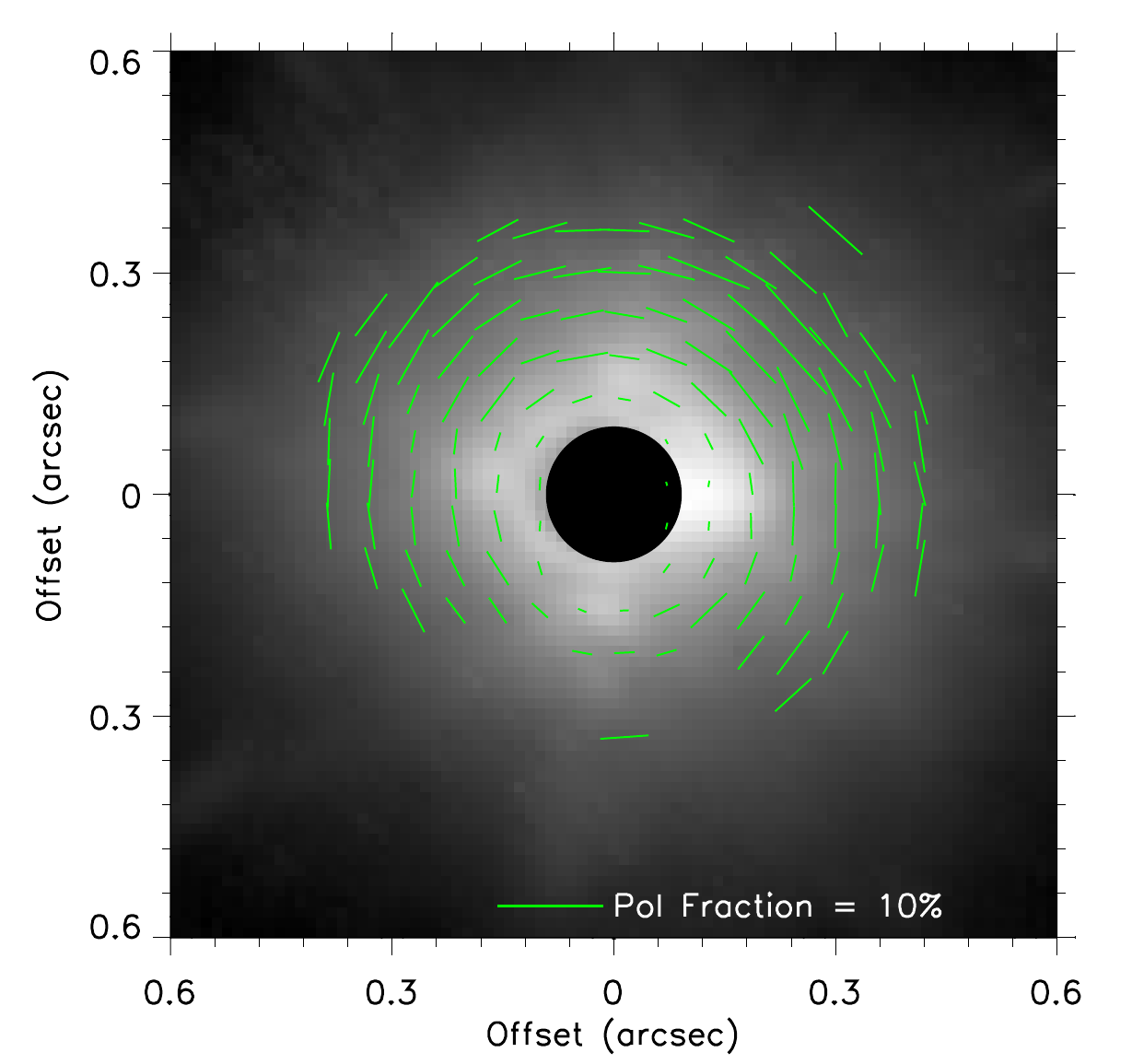} 
\includegraphics[width=2in]{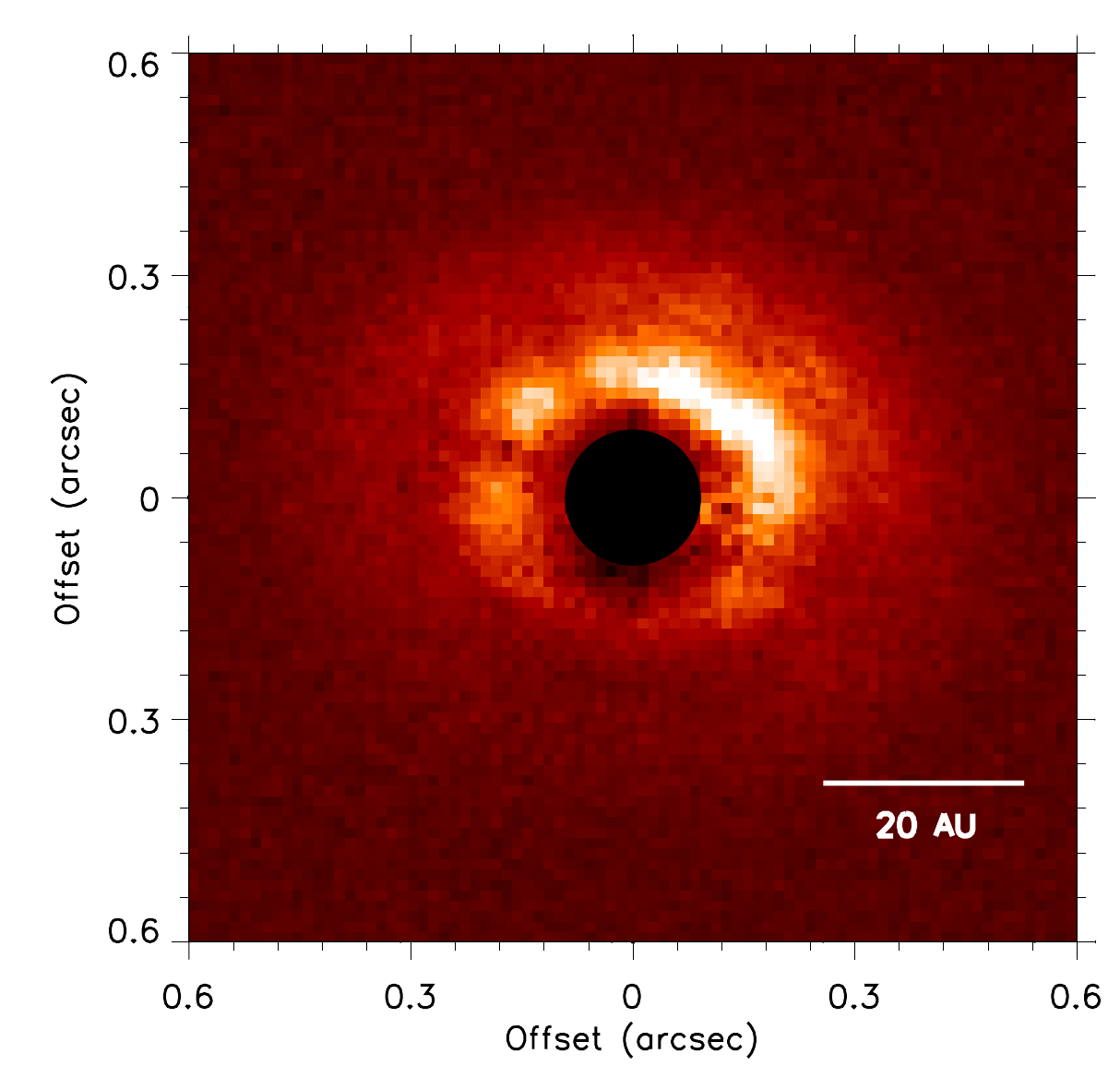} 
\includegraphics[width=2.15in]{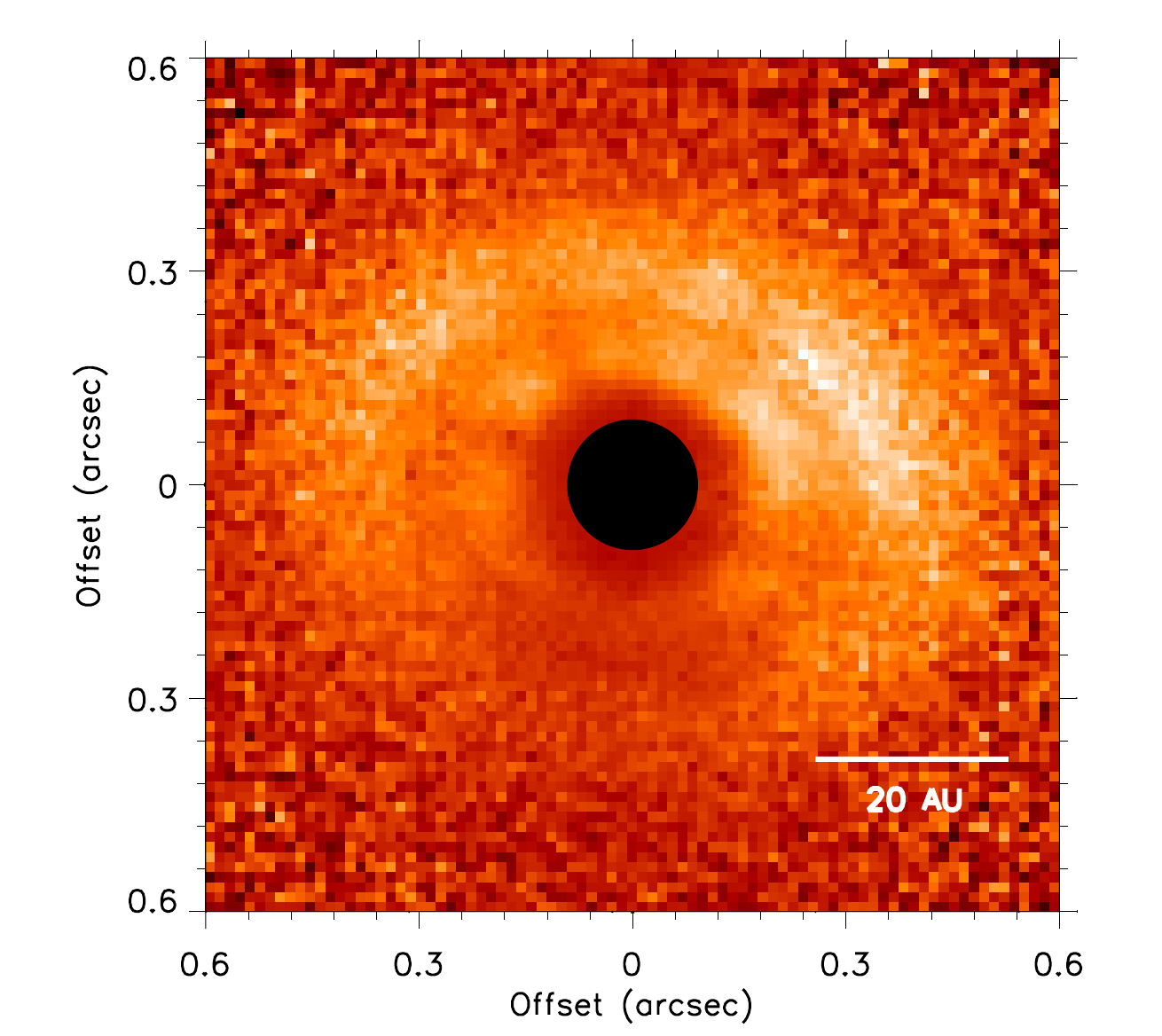} 
\includegraphics[width=2in]{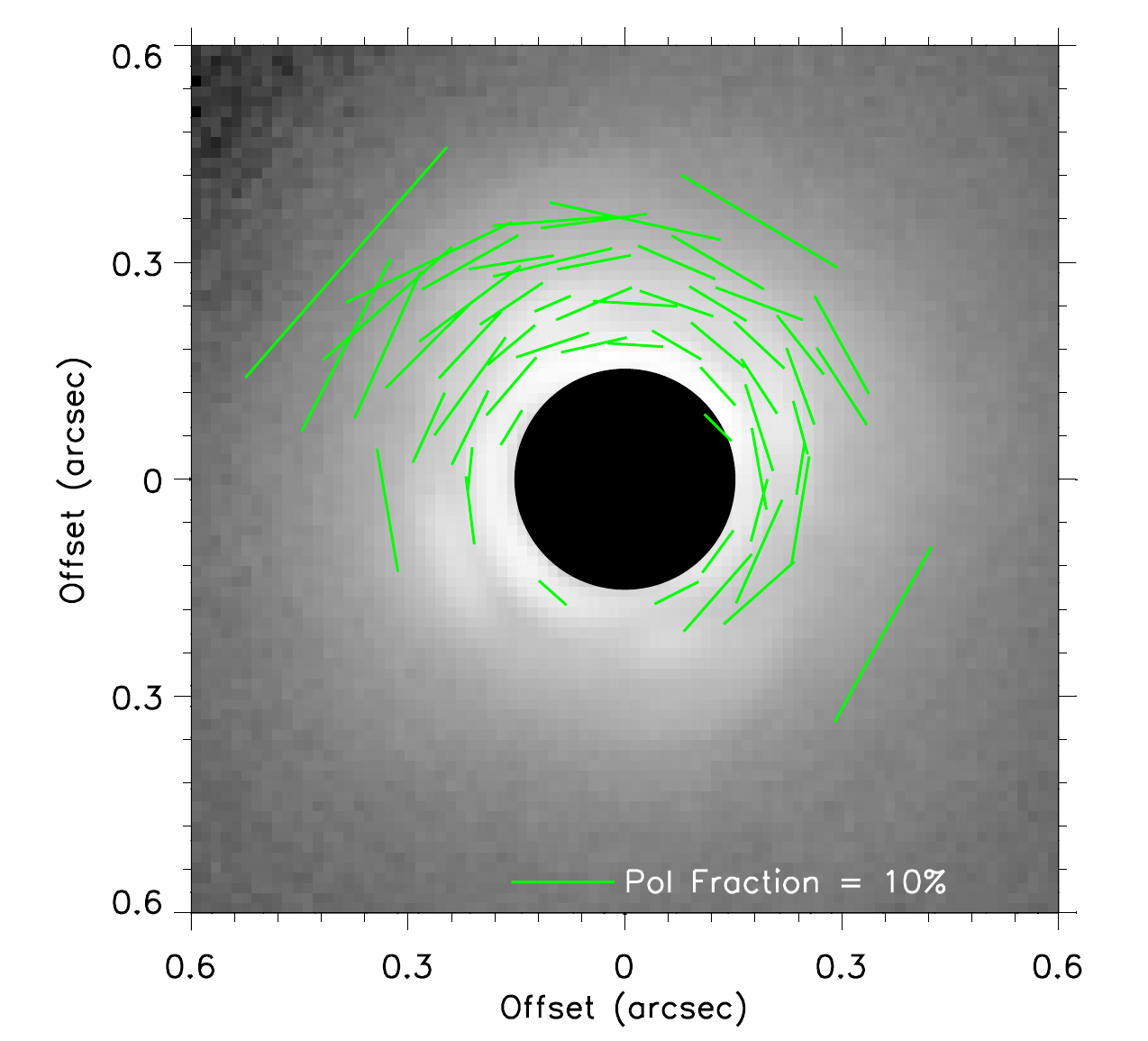} 
\includegraphics[width=2in]{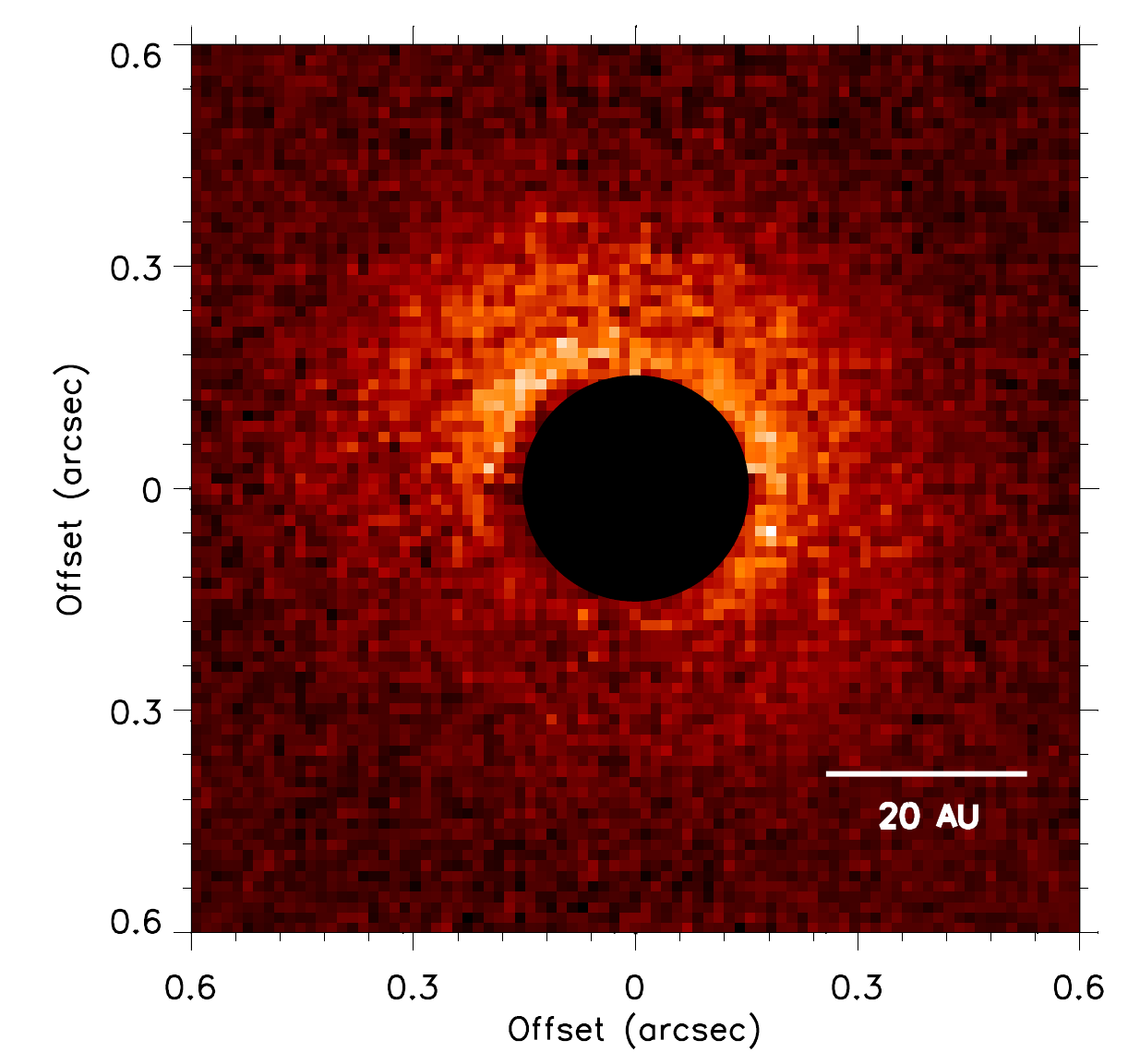} 
\includegraphics[width=2.1in]{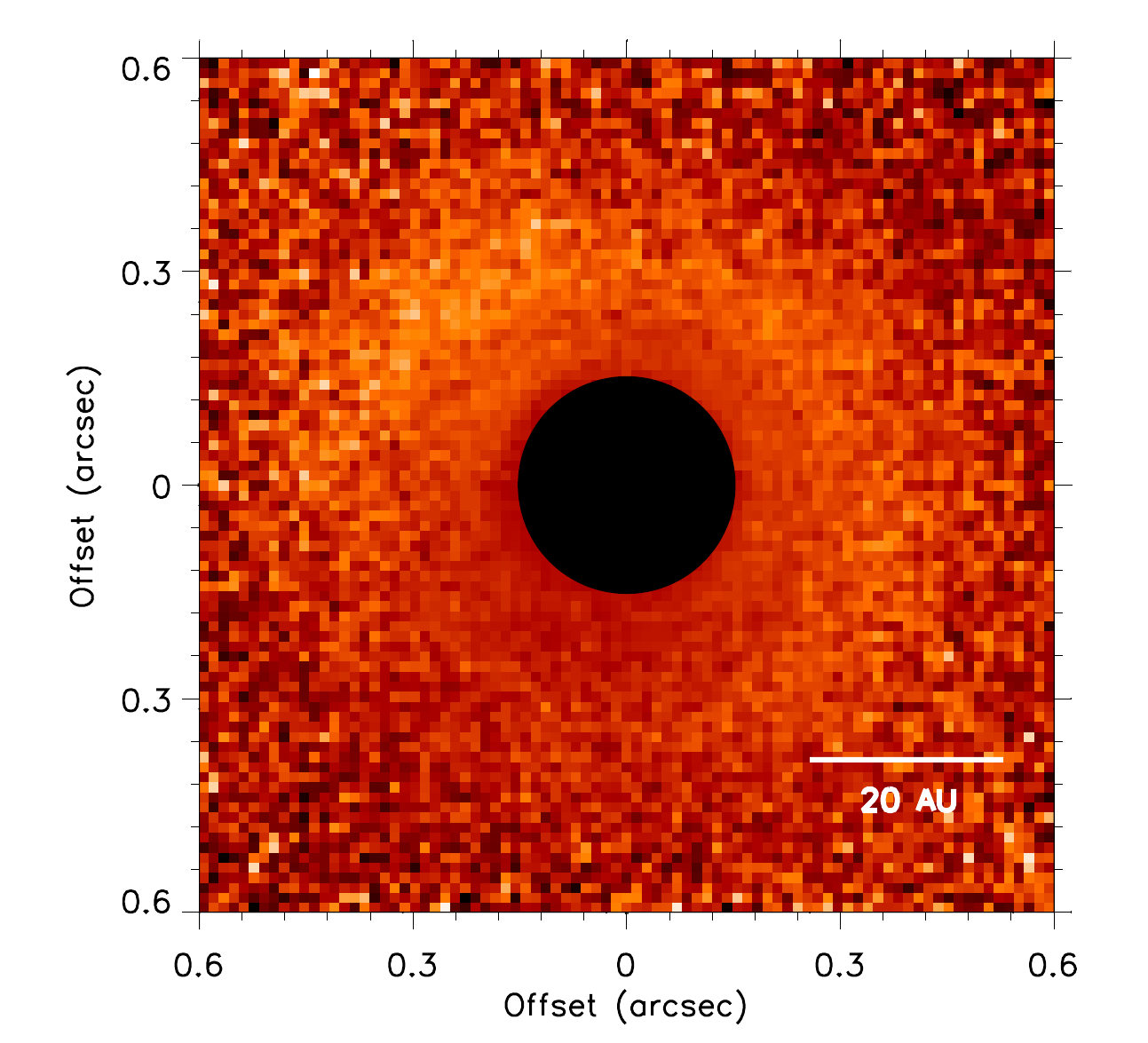} 

\caption{Left : Total intensity J (top) and K2 (bottom) images with polarization degree (p=P/I) (pseudo-)vectors overlaid in green for pixels where the total polarized intensity is greater than 40 counts. Middle : J (top) and K2 (bottom) polarized intensity (P$_\perp$) images. Right: P$_\perp$ scaled by r$^{2}$, where r is the distance in pixels from the central binary, corrected for projection effects. All images are shown on a linear scale. The coronagraph is represented by the black filled circles and images are oriented with north up and east to the left. A small artifact from slight telescope mispointing during acquisition of the J-band image sequence can be seen to the west of the coronagraph in the P$_\perp$ images.}
\end{center}
\end{figure}

The ring structures seen in the P$_{\perp}$ images cannot be attributed to the point spread function of the occulted central star system, given that similarly bright stars imaged by GPI in coronagraphic/polarimetric mode show no such features \citep[e.g.][]{Perrin2014}. Furthermore, the inclination, position angle, and north/south brightness contrast of the disk dust ring system are consistent with the outer ($\sim$29-350 AU) disk orientation and inclination as inferred from interferometry of mm-wave CO emission. Given that circumstellar dust grains in disks orbiting young stars preferentially scatter starlight in the forward direction \citep[e.g.][]{Perrin2014}, we conclude that the north side of the disk is tipped toward Earth, consistent with the disk orientation inferred from Submillimeter Array (SMA) $^{12}$CO observations of V4046 Sgr \citep{Rosenfeld2012}. 

To better characterize the double ring structure that is evident in the polarized intensity images, we constructed radial brightness profiles from the J band image in the East-West and South-North directions, after rotating to account for the disk equatorial plane position angle of 76$^{\circ}$ (Fig. 2). Both radial profiles show a distinct inner ring that peaks in brightness at $\sim$14 AU, surrounded by a "shoulder" and a fainter halo extending out to $\sim$45 AU. The inner ring, which has a FWHM of $\sim$10 AU (as measured using the eastern-directed radial profile in Fig. 2), blends smoothly with the outer halo in the North, but a distinct break is evident at $\sim$20 AU in the East, West and South.

\begin{figure}
\begin{center}
\includegraphics[scale=0.75]{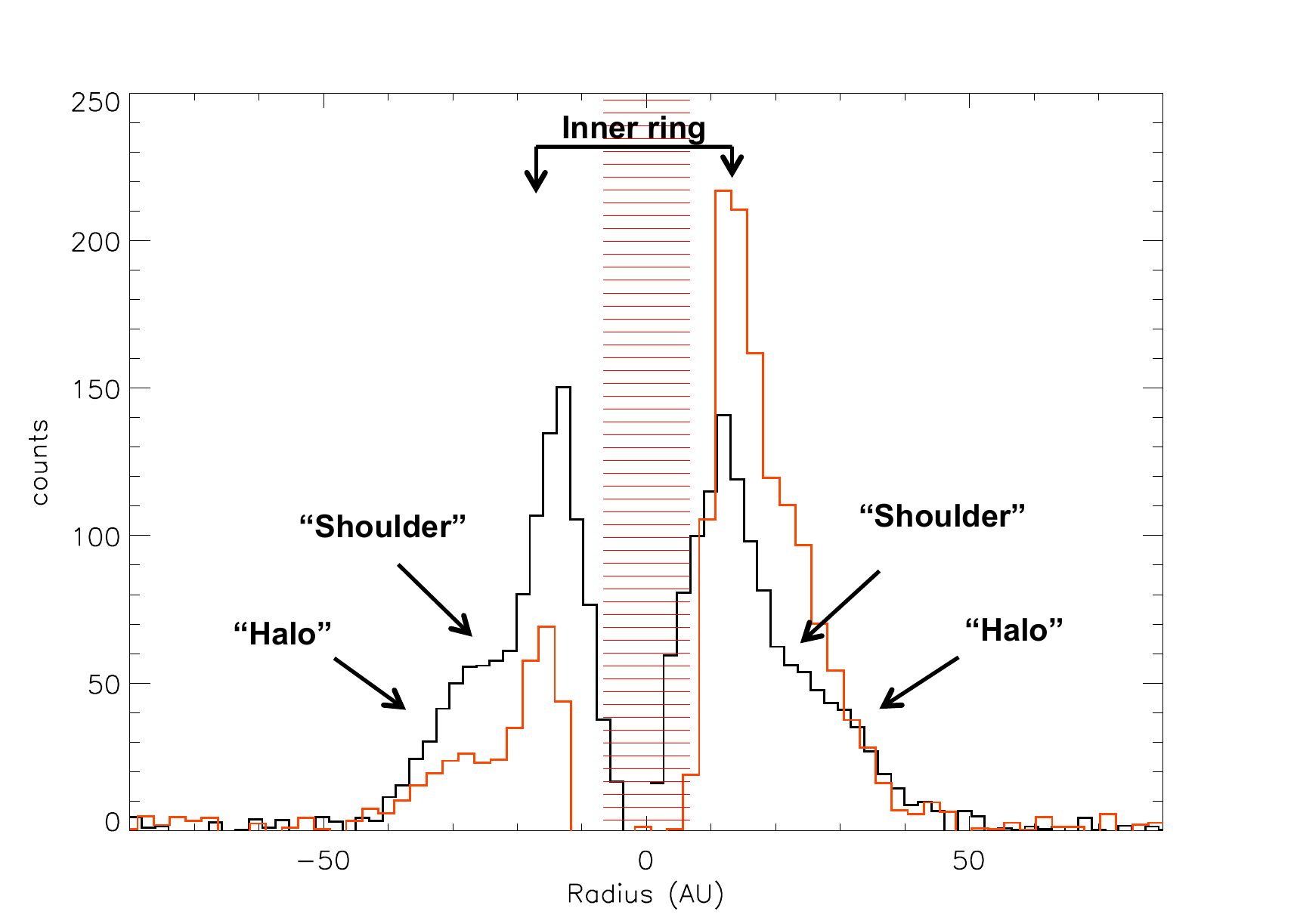}
\caption{Radial profile extracted from the J band P$_{\perp}$ image binned 2x2 along the 76$^{\circ}$ position angle showing the East-West (black) and South-North (orange) brightness profile of the disk. The horizontal red lines show the location of the coronagraph. The uncertainties in each bin of the P$_\perp$ profiles, determined from P$_{\parallel}$, at both J and K2 is $\sim$6 counts.}
\end{center}
\end{figure}

In Figure 3 we present surface brightness profiles at J and K2 created by averaging concentric elliptical annuli at 0.014" intervals (the angular size of one pixel) with minor:major axis ratios of 0.84 (approximating a circular disk with inclination of 33.5$^\circ$) oriented at a position angle of 76$^\circ$ \citep{Rosenfeld2012}. Both the J and K2 surface brightness profiles display a roughly power-law dependence on radius (i.e., surface brightness $\propto r^{-n}$) between $\sim$14 AU and $\sim$45 AU, which is the apparent outer limit of scattered light detected in the radial profiles. In both bands, there is a clear break in the power-law index at $\sim$28 AU, wherein $n \approx 2.0$ in the range $\sim$14 - 28 AU and $n \approx 5.5$ at larger radii. The former result is consistent with simple geometrical dilution of starlight. 

\begin{figure}[h]
\begin{center}
\includegraphics[scale=0.4]{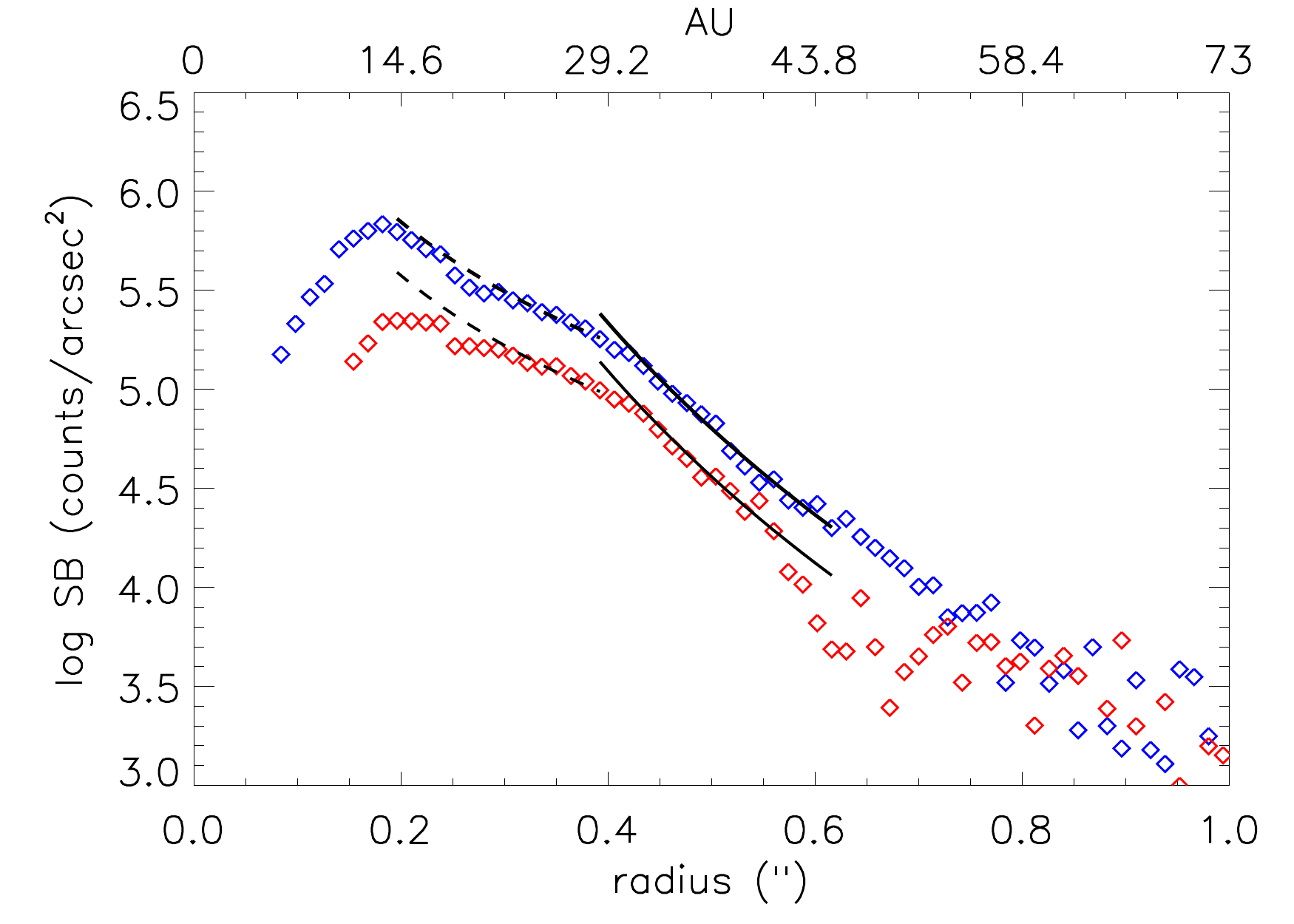}
\includegraphics[scale=0.4]{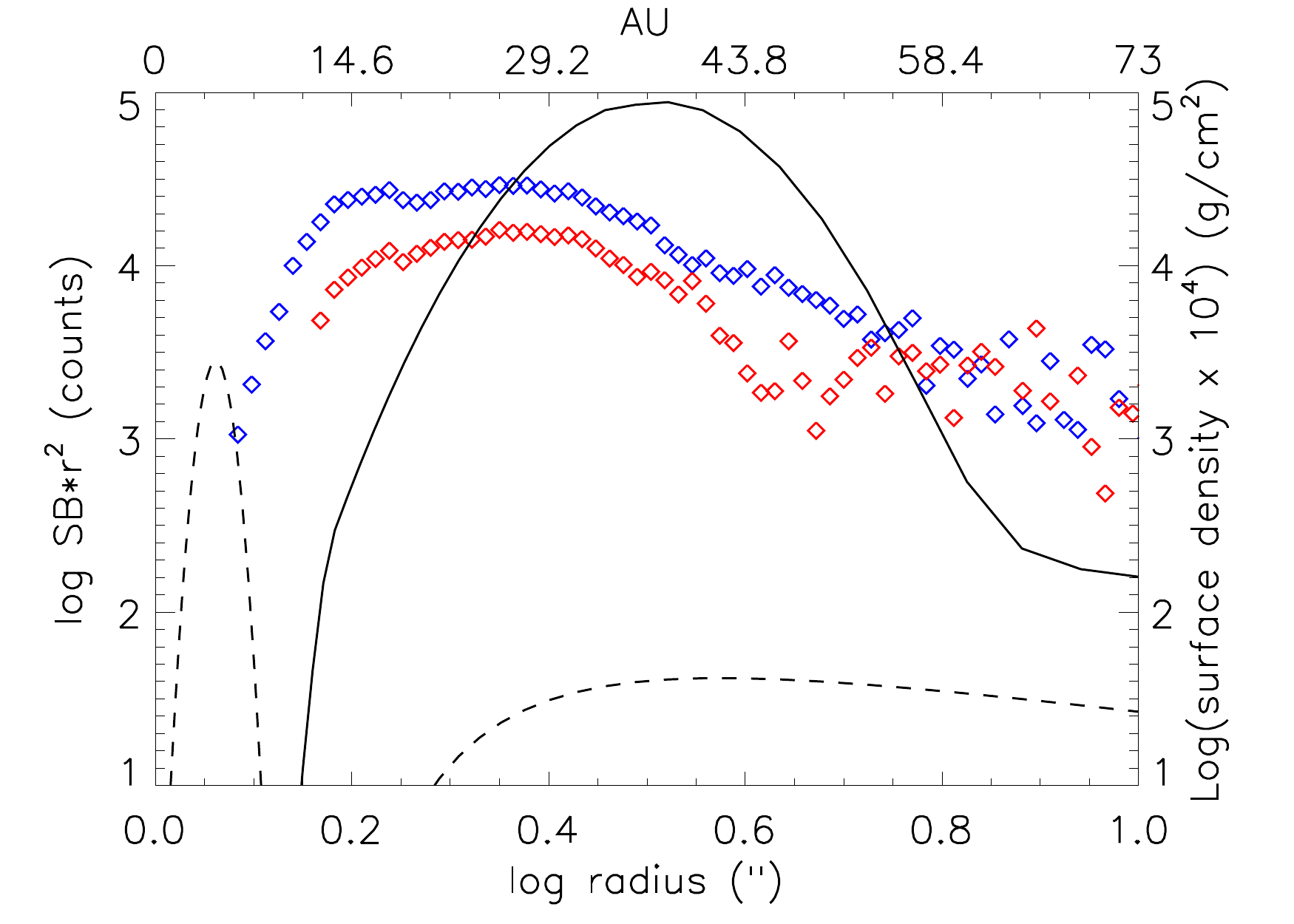}
\caption{Left:  J (blue) and K2 (red) surface brightness (SB) curves. The error bars are smaller than the symbol size. The black dashed line represents r$^{-2}$ fit from $\sim$14 - 28 AU and black solid line represents r$^{-5.5}$ fit from $\sim$28 - 45 AU. Right: Background subtracted surface brightness curves multiplied by r$^{2}$ with the surface density of small ($\mu$m-sized; black dashed) and large (mm-sized; black solid) dust grains from the \citet{Rosenfeld2013} model overlaid. The surface density of both the small and large dust grains has been scaled up by 10$^{4}$.}
\end{center}
\end{figure}

\section{Discussion}

\subsection{Comparison of GPI and SMA imaging of V4046 Sgr}

\citet{Rosenfeld2013} modeled 1.3 mm CO and continuum interferometric imaging data for V4046 Sgr in terms of a ring of large (mm-sized) dust grains following a Gaussian ring surface density profile with a mean radius of 37 AU and FWHM of 16 AU, surrounded by an extended ($\sim$350 AU radius) halo of molecular gas and small grains. In this model, the region interior to $\sim$29 AU is depleted of large grains, but contains smaller ($\mu$m-sized) dust. This model also well reproduces mid- to far-infrared spectrophotometry of V4046 Sgr (Rapson et al. 2014, submitted). Figure 4 compares the GPI total linear polarized intensity with the SMA 1.3mm continuum emission map \citep{Rosenfeld2013}. The fainter, outer scattered light halo in the GPI polarized intensity images, which extends to $\sim$45 AU, appears to merge into the peaks in SMA-detected flux. Figure 3 (right) shows the surface brightness profiles at J and K2 with the surface density of small ($\mu$m-sized) and large (mm-sized) dust grains from the \citet{Rosenfeld2013} model overlaid. We clearly see that the dust in our GPI images fills the gap interior to $\sim$29 AU, and extends into the ring of mm-sized dust that peaks at 37 AU. The \citet{Rosenfeld2013} model also predicts the existence of $\mu$m-sized grains interior to $\sim$7 AU, a region that lies behind the coronagraph in the GPI images. Evidently, the inner dust ring imaged by GPI has no counterpart in either the small- or large-grain model components modeled by \citet{Rosenfeld2013}.

\begin{figure}[ht]
\begin{center}
\includegraphics[width=6in]{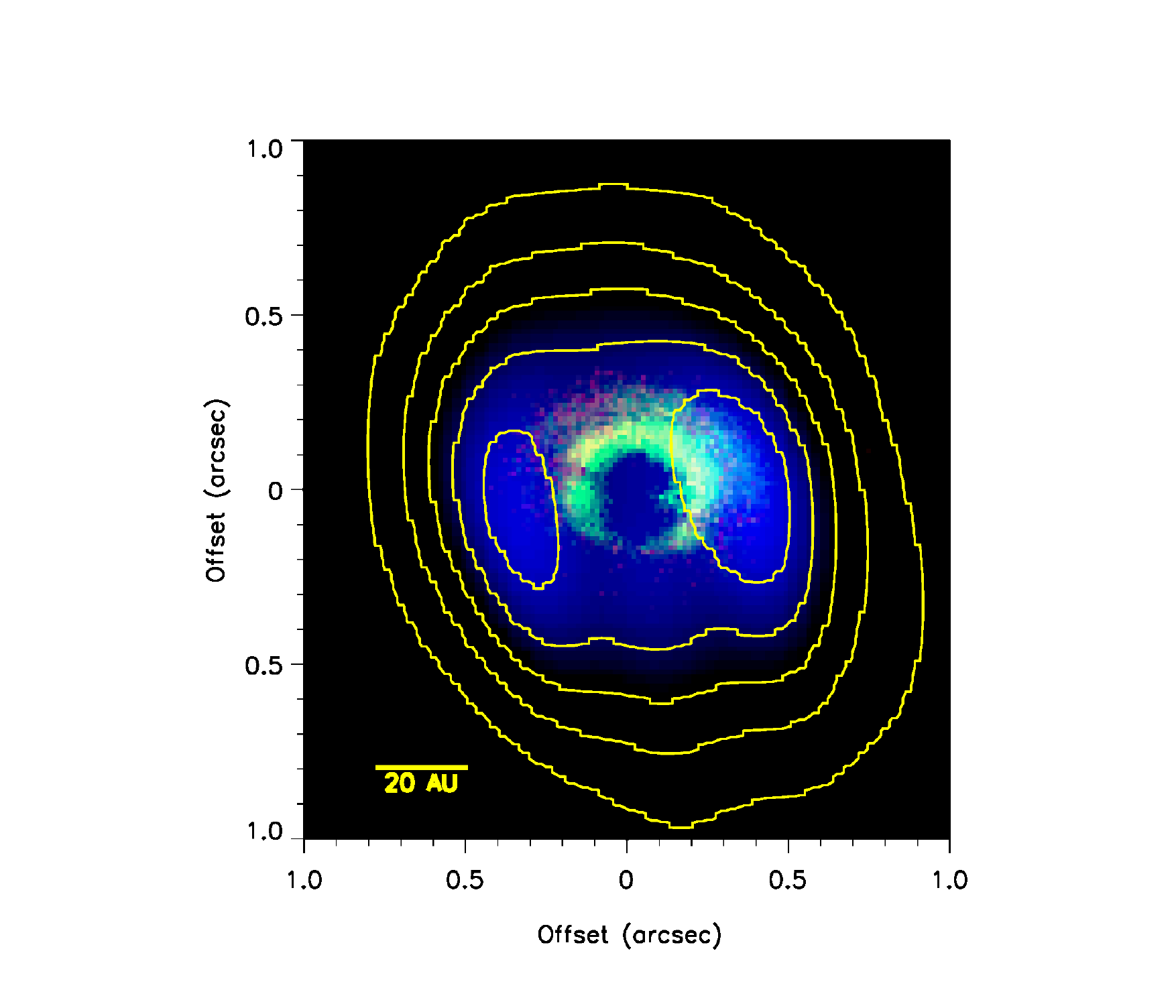}
\caption{ Three color composite image comparing SMA 1.3mm continuum emission \citep[blue with yellow contours overlaid;][]{Rosenfeld2013} and GPI J (green) and K2 (red) total linear polarized intensity. The SMA data has a beam size of $0.74" \times 0. 38"$ and has been registered such that the peaks of the mm-wave continuum emission lie equidistant from the coronagraph center.}
\end{center}
\end{figure}
\pagebreak

\subsection{Evidence for radial dust segregation by size}

The foregoing comparison between GPI and SMA imaging demonstrates that the inner $\sim$45 AU of the disk has undergone significant dust particle growth and particle size segregation. Such particle growth and migration processes are expected to accompany an epoch of giant planet formation. Specifically, theories of giant planet formation in dusty disks with embedded, nascent planets predict the generation of radially localized pressure maxima as a consequence of planet-disk dynamical interactions \citep{Rice2006,Zhu2012}.  These pressure gradients trap larger (mm- to cm-sized) particles outside the planet-forming regions of the disk, whereas smaller (micron-sized) grains freely pass through the pressure traps, resulting in strong dust particle size gradients.

Interferometric imaging of dusty protoplanetary disks often reveal central disk clearings whose inner radii are significantly larger than the orbits of the Jovian planets in our solar system \citep[e.g.][]{Andrews2011}. This dichotomy in disk size scales has left open the potential connection between inner disk clearings and giant planet building. Modeling by \citet{Pinilla2012} shows that pressure bumps in the disk due to gaps opened by orbiting planets trap mm-sized dust grains into a ring that may be located at radii greater than twice a planet's orbital radius, depending on the mass of the planet. Smaller ($\lesssim \mu$m-sized) dust particles drift inward and create a ring between the planet-induced gap and the mm dust ring.
 
Such an interpretation has been applied to mm-wave and near-IR imaging of the disk orbiting the intermediate-mass star SAO 206462 \citep{Garufi2013}. The surface brightness profile from near-infrared scattered light off the disk around SAO 206462 shows a similar power law dependence on scales about twice that determined here for V4046 Sgr, despite the fact that the SAO 206462 disk exhibits spiral structure, rather than rings, and the scattered light lies exterior to the sub-mm hole. Thus, the dust ring structure and radial profiles seen in both V4046 Sgr and SAO 206462 suggest that dust segregation and planet formation may be occurring within these disks.

\subsection{Implications for planet formation in the disk around V4046 Sgr}

It is possible that the depletion in scattered light at R$\lesssim$ 14 AU in our GPI images may be due to giant planet formation. This cavity cannot be due purely to dynamical effects from the binary system, as the tight ($\sim$9 $R_\odot$) binary can only truncate the disk out to $\sim$0.135 AU. Although this cavity could conceivably be filled by a geometrically thin disk surface that happens to lie roughly parallel to the paths of incoming photons from the central stars \citep[e.g.][]{Takami2014}, we regard such an ({\it ad hoc}) inner disk geometry as less likely than that of an opening carved out by a giant planet. It is also possible, but unlikely, that photoevaporation has caused this depletion of dust within $\sim$14 AU, since models predict that micron-sized dust is rapidly destroyed by X-ray/UV radiation \citep{Gorti2009,Owen2012}, and the GPI images demonstrate that a significant mass of micron-sized dust is still present at disk radii between 14 to 45 AU after $\sim$20 Myr of disk evolution.

 \citet{Ruge2014} model the effects of giant planets in massive, dusty disks on the appearance of gaps in scattered light images. They find that the formation of a giant planet results in a deficit of 1.3 mm emission at the location of the forming planet for disks more massive than 2.67$\times$10$^{-3}$ M$_{\odot}$. Gaps in near-infrared scattered light are also evident for lower mass disks, but for massive disks (M $\gtrsim$2.67$\times$10$^{-3}$ M$_{\odot}$) the disk remains optically thick and thus a gap may not be apparent in scattered light at near-infrared wavelengths. While the V4046 Sgr disk is massive overall (M$\sim$0.1M$_{\odot}$), modeling of the disk dust distribution \citep{Rosenfeld2013}, and the evidence for dust segregation by size, suggests most of the dust mass resides beyond the $\sim$29 AU sub-mm ``edge", thereby allowing for gaps formed by giant planets to be visible in scattered light images at near-infrared wavelengths. These models suggest that planets may be forming interior to $\sim$14 AU and possibly near the R$\sim$20 AU break between the rings seen in scattered light (Figs. 2,3). Considering the depletion of sub-mm emission interior to $\sim$29 AU \citep{Rosenfeld2013}, a giant planet at $\sim$20 AU would also be consistent with modeling predicting that sub-mm rings form at $\gtrsim$1.5 times the planets orbital radius \citep{Rice2006, Paardekooper2004}. 

Simulations of planet formation in circumstellar disks by \citet{Dong2014} further support the idea that planet formation is occurring in the V4046 Sgr disk. These models show that Jupiter-mass planets can clear large visible gaps in $\mu$m and mm wavelengths at the locations of forming planets. Planets interior to a sub-mm cavity can form multiple dust rings and gaps depending on the planet's size and location, just as we observe in our GPI images of V4046 Sgr (compare our Figs. 1 and 3 with their Figs. 2,5 and 7). Overall, our GPI imaging of the V4046 Sgr disk hence provides vivid evidence in support of so-called ``dust filtration'' models describing the structure of protoplanetary disks following giant planet formation. 

\section{Conclusions}

GPI coronagraphic/polarimetric imaging of the V4046 Sgr AB circumbinary disk, combined with mm imaging, demonstrates dust segregation by size into rings, which may be caused by multiple young giant planets orbiting the V4046 Sgr  AB binary system at orbital semimajor axes similar to those of the giant planets in our solar system. Polarized intensity images yield evidence that the disk is well populated by relatively small (micron-sized or smaller) grains within its central, $\sim$29 AU large-grain cavity and, furthermore, that there exists an interior $\lesssim$14 AU-radius region that is devoid even of small grains. Comparison with models of protoplanet-disk interactions \citep{Pinilla2012,Ruge2014,Dong2014} suggests that gas giant planets may be present, and actively carving gaps, at R$\lesssim$14 AU and at R$\sim$20 AU in the V4046 Sgr disk.

Further imaging with Gemini/GPI or similar instrumentation at higher signal to noise, as well as high resolution imaging with ALMA, is necessary to better discern the structure of the rings evident around V4046 Sgr. Modeling is also needed to determine the grain properties and dust mass within the disk, along with the location of possible forming planets. Further investigations aimed at directly and indirectly detecting potential young giant planets orbiting V4046 Sgr AB will also provide essential constraints on simulations aimed at understanding the conditions in which giant planets might form in circumbinary orbits --- a theoretical question that is presently of intense interest, given the Kepler Mission's detection of circumbinary planets \citep{Pierens2013}.

\section*{Acknowledgements}
This work is based on observations obtained at the Gemini Observatory, which is operated by the Association of Universities for Research in Astronomy, Inc., under a cooperative agreement with the NSF on behalf of the Gemini partnership: the National Science Foundation (United States), the National Research Council (Canada), CONICYT (Chile), the Australian Research Council (Australia), Minist\'{e}rio da Ci\^{e}ncia, Tecnologia e Inova\c{c}\~{a}o (Brazil) and Ministerio de Ciencia, Tecnolog\'{i}a e Innovaci\'{o}n Productiva (Argentina). Support is provided by the National Science Foundation grant AST-1108950 to RIT.

\bibliographystyle{apj}


\end{document}